\documentclass[10pt,letterpaper]{article}
\usepackage{opex3}
\usepackage{cite}

\newcommand*{\XYZ}{}

\begin{document}
\title{Femtosecond-scale switching based on excited free-carriers}

\author{Y. Sivan$^{1,*}$, G. Ctistis$^{2,3,4}$, E. Y\"uce$^2$ and A.P. Mosk$^2$}
\address{$^1$ Unit of Electro-Optics Engineering, Faculty of Engineering Sciences, Ben-Gurion University of the Negev, Beer-Sheva, Israel, P.O. Box 654, 84105.

$^2$ Complex Photonic Systems (COPS), MESA+ Institute for Nanotechnology, University of Twente, P.O. Box 217, 7500 AE Enschede, The Netherlands.

$^3$ NanoBioInterface Research Group, School of Life Science, Engineering, and Design, Saxion University of Applied Sciences, M.H. Tromplaan 28, P.O. Box 70.000, 7500 KB Enschede, The Netherlands.

$^4$ Integrated Optical Microsystems Group, MESA+ Institute for Nanotechnology, University of Twente, P.O.Box 217, 7500 AE Enschede, The Netherlands.}

\email{*sivanyon@bgu.ac.il} 

\begin{abstract}
  We describe novel optical switching schemes operating at femtosecond time scales by employing free carrier (FC) excitation. Such unprecedented switching times are made possible by spatially patterning the density of the excited FCs. In the first realization, we rely on diffusion, i.e., on the nonlocality of the FC nonlinear response of the semiconductor, to erase the initial FC pattern and, thereby, eliminate the reflectivity of the system. In the second realization, we erase the FC pattern by launching a second pump pulse at a controlled delay. We discuss the advantages and limitations of the proposed approaches and demonstrate their potential applicability for switching ultrashort pulses propagating in silicon waveguides. We show switching efficiencies of up to $50\%$ for $100$ fs pump pulses, which is an unusually high level of efficiency for such a short interaction time, a result of the use of the strong FC nonlinearity. Due to limitations of saturation and pattern effects, these schemes can be employed for switching applications that require femtosecond features but standard repetition rates. Such applications include switching of ultrashort pulses, femtosecond spectroscopy (gating), time-reversal of short pulses for aberration compensation, and many more. This approach is also the starting point for ultrafast amplitude modulations and a new route toward the spatio-temporal shaping of short optical pulses.
\end{abstract}

\ocis{(190.2055) Dynamic gratings; (250.4110) Modulators; (190.5530) Pulse propagation and temporal solitons; (190.5910) Semiconductor nonlinear optics; (320.7130) Ultrafast processes in condensed matter, including semiconductors}.



\section{Introduction}

Developing the ability to control the optical properties of matter dynamically has been a topic of intense research for many decades. 
Such control is necessary, for instance, in modern computer technology, which is based largely on semiconductor components - such as transistors and diodes - that are the building blocks of analog and digital circuits. The functionality of such components is based on the ability to modify their performance reversibly at a high bit rate. Similarly, in communication systems, {\em switchable} semiconductor components are used for tasks such as information encoding, modulating, multiplexing, routing, and frequency-shifting.

{\XYZ In these cases, modifications of the refractive index are typically performed repetitively within fixed time intervals that are rarely shorter than $\approx 10$ ps, i.e., at sub-THz rates.
Many other applications, e.g., time-resolved spectroscopy and optical data processing, require faster modifications (i.e., having femtosecond features), yet with no need to exceed the THz repetition rate~\cite{Eggleton_OPN,Emre_OL_2013}. Fundamental studies, such as real-time cavity quantum electrodynamics and the control of quantum states of matter, likewise require controlling the refractive index at femtosecond time scales. In the current work,
for brevity, we use the term ``switching'' to describe a general modification of the refractive index that occurs repetitively at a given rate, but the modification does not need to be purely sinusoidal. 
In some places, this process is also referred to as the dynamic tuning of the refractive index~\cite{Fan-OPN,Fan-reversal,Fan-stopping,Longhi-reversal,Sivan-Pendry-letter,Sivan-Pendry-article}. }

The main mechanism in use for modifying the optical properties of solids is the excitation of free charge-carriers  (FCs)~\cite{Vos_PRB_2002,Wehrspohn_PRB_2002,Gaeta_PRB_2004,Euser_Vos_JAP_2005,Lipson_review,Vos_APL_2007,Euser_Thesis}. FC excitation is frequently accomplished {\em electrically}, e.g., by an injection or a depletion of carriers with a bias voltage. However, there is also growing interest in {\em optical} switching, whereby a strong pump pulse with a frequency that exceeds the energy gap is absorbed and generates electron-hole (e-h) pairs~\cite{Euser_Thesis,Euser_Vos_JAP_2005,Lipson_review,Othonos_review}. 
This plasma has a Drude-like contribution to the dielectric constant of the semiconductor, resulting in a refractive index that is lowered proportionally to the pump intensity (i.e., it is a nonlinear effect), reaching values as high as $\Delta n / n \sim 10^{-1}$~\cite{Lipson_review,Euser_Thesis}. However, while the decrease of the refractive index occurs at time scales ranging from a few tens of femtoseconds (e.g., in silicon) to several picoseconds (e.g., in GaAs)~\cite{Euser_Thesis}, the refractive index relaxes to its equilibrium value at much longer time scales, depending on the dominant material relaxation mechanisms. These range from no less than several tens of picoseconds to a few nano-seconds~\cite{van_Driel_kinetics_1987,Riffe_Si_e_kinetics_2002,Othonos_review}.

A faster alternative for {\em optical} switching employs the Kerr effect, a nonlinear optical effect originating from the non-resonant response of bound, low-mass electrons and results in nearly instantaneous response times. However, the changes to the refractive index associated with the Kerr effect are weak: they rarely exceed $\Delta n / n \sim 10^{-5} - 10^{-4}$~\cite{Euser_Thesis,Boyd-book}. Accordingly, the efficiency in Kerr switching is low even for very high intensities.

To benefit from the best of both worlds, i.e., to achieve fast yet large refractive index changes, several routes have been explored with the aim of shortening the lifetime of the FC. Popular approaches include resonant phonon absorption~\cite{Gerard_GaAs_mid_IR_relaxation}, reverse-bias on semiconductor heterostructures~\cite{Lipson_ps_lifetime_Si_wgs,Lipson_PINIP,Feldman,LiKamWa}, ion implantation~\cite{chi_ultrafast_ion-damages_si,Paulter_ultrafast_ion-damages_gaas,ion-implant-Silberberg_1985,ps_Si_modulation,ultrafast_H_plus_InP_APL_1991}, or operating at low temperatures~\cite{Mourou_ultrafast_low_T,Fauchet_ultrafast_low_T,Elezzabi_ultrafast_low_T}. However, such approaches can reduce the FC lifetime to a few picoseconds, but not faster. Worse, this is usually achieved at the cost of increased absorption and scattering and of mounting technological complexity.

{\XYZ In the current paper, we propose two alternative approaches for femtosecond-scale FC-based switching Unlike standard optical switching, which predominantly employs a single intense pump pulse that is {\em spatially uniform}, we consider a configuration based on a {\em highly spatially non-uniform} pump pulse. In particular, we consider the case where the index modification varies periodically in one spatial direction. Such configurations are known as Bragg gratings (BGs) and} are the simplest periodic photonic structures, also known, by analogy with solid-state physics, as photonic crystals~\cite{Joannopoulos-book}. The periodicity of the optical properties creates a spectral band in which the propagation of electromagnetic waves perpendicular to the grating is prohibited. The BGs induced by the pump have a finite temporal duration ({\em Transient} Bragg gratings, TBGs) and are, in effect, a sub-class of dynamically-tuned photonic crystals~\cite{Fan-OPN}. TBGs were proposed for various applications, including as wavelength filters and multiplexers in optical communication systems, as highly reflective mirrors in fiber lasers, or as temperature and strain gauges in fibers. On the fundamental level, TBGs have been used for spectroscopy in gases and for studying FC recombination rates in semiconductors, see~\cite{Othonos_review} and references therein. {\XYZ They can be generated through interference of two pump beams, by using phase masks, or through the generation of a (transient) standing wave profile, as in photosensitive fibers~\cite{photosensitive-fiber-BGs-original}.}

\begin{figure}[htbp]
  \centering{\includegraphics[scale=0.65]{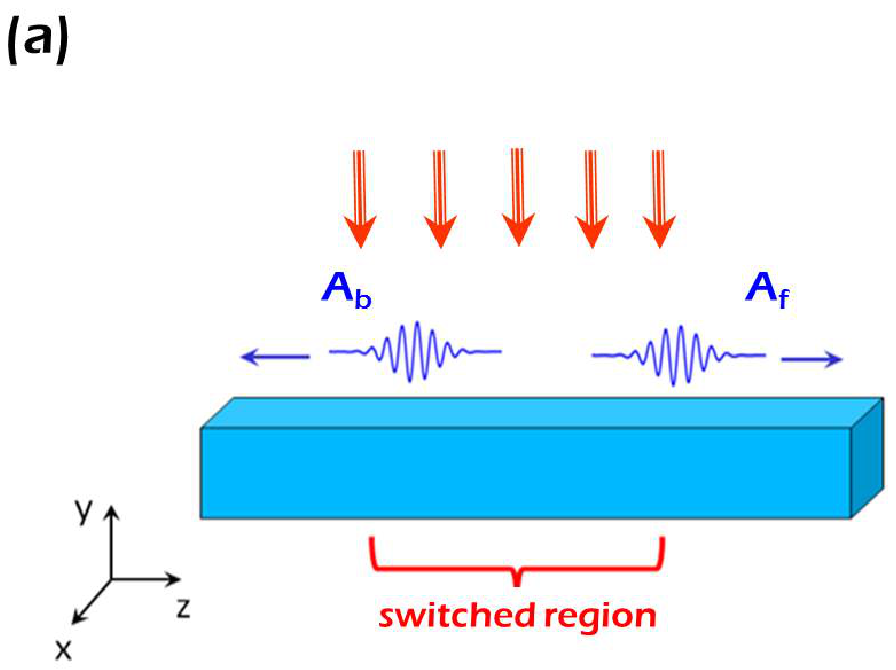}\includegraphics[scale=0.65]{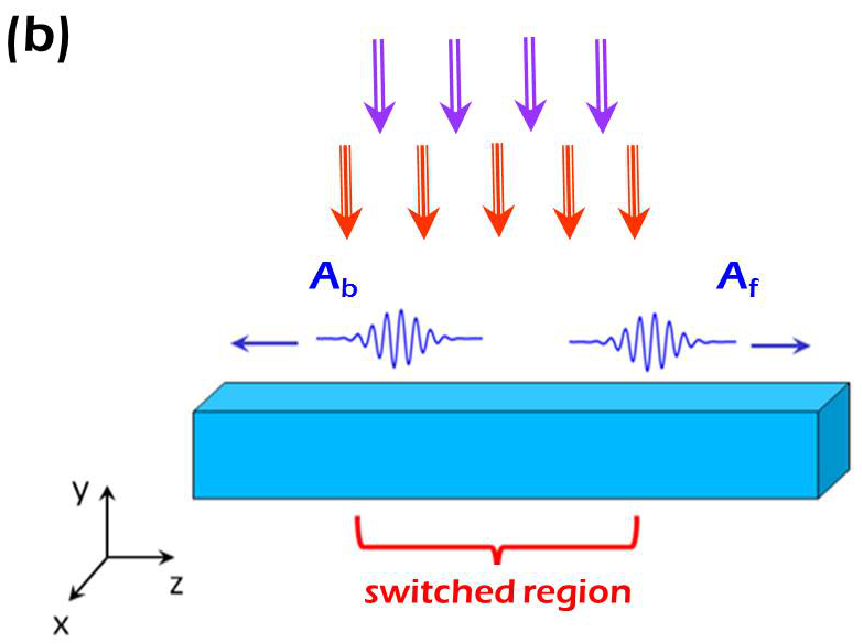}}
  \caption[]{Schematic illustration of the proposed switching techniques. (a) A single periodically modulated pump (red arrows; note that the distance between the arrows is assumed to correspond to half the wavelength of the forward signal) which generates a pattern that self-erases due to diffusion. (b) A dual pump scheme, in which the second pump pulse is shifted by half the spatial period, such that it erases the index contrast generated by the first pump pulse.
   } \label{fig:schematics}
\end{figure}

In our configuration, we utilize the unique ability of TBGs to be erased on almost arbitrarily fast times scales (Fig.~\ref{fig:schematics}). 
In one realization (see Fig.~\ref{fig:schematics}(a)), we rely on an intrinsic property of the FCs, namely, diffusion, to induce {\em self-}erasure of the BG; in the second realization (see Fig.~\ref{fig:schematics}(b)), we rely on a slightly more complex switching scheme consisting of two spatially staggered pump pulses. By using these approaches, we demonstrate {\em theoretically} switching {\em times} that are as short as a few tens of femtoseconds, i.e., up to 2 orders of magnitude shorter than the fastest standard switching times in semiconductors. {\XYZ Importantly, this decrease in switching times is achieved just with a simple modification of the illumination scheme, hence, without the increased absorption and scattering associated with the standard material-based approaches~\cite{Gerard_GaAs_mid_IR_relaxation,chi_ultrafast_ion-damages_si,Paulter_ultrafast_ion-damages_gaas,ion-implant-Silberberg_1985,ps_Si_modulation,ultrafast_H_plus_InP_APL_1991}, and with no need to resort to low temperatures~\cite{Mourou_ultrafast_low_T,Fauchet_ultrafast_low_T,Elezzabi_ultrafast_low_T}. }

To achieve the fastest response, we focus on switching low quality (Q)-factor systems~\cite{Vos_PRB_2002,Euser_Vos_JAP_2005,Vos_APL_2007,ctistis_APL_2011_Kerr}. Although we use the terminology and schematics appropriate for {\em optical} switching of semiconductors, the concept also applies to more general (e.g., {\em electronic-based}) switching, e.g., with interleaved PN junctions~\cite{Interleaved_pns_Yu_OE_2009,Interleaved_pns_Yu_OE_2012}, and to other material platforms. {\XYZ In some of these cases, one may need to distinguish between majority and minority carriers.}

The paper is organized as follows. In Section~\ref{sec:FC-nlties}, we review the standard model for FC nonlinearity and show how the limitations of standard switching approaches are manifested in this model. In Section~\ref{sec:diffusion-switch}, we introduce the diffusion-based switching approach, which relies on exciting an initial periodic pattern of FCs. In Section~\ref{sec:we-switch}, we introduce an even faster switching technique that relies on a double pump scheme, in which the second pump erases the periodic pattern generated by the first pump. In Section~\ref{sec:model}, we introduce a simple model for pulse propagation in a semiconductor waveguide based on coupled-mode theory~\cite{Yeh-book}. In Section~\ref{sec:model-solution}, we provide specific examples for the two novel switching schemes, based on numerical solutions of the model, and discuss their implementation for time-reversal of short optical pulses. In Section~\ref{sec:discussion}, we summarize the results, compare them to previous works and provide a brief outlook for possible future extensions.

\section{Free-carrier nonlinearities}\label{sec:FC-nlties}

The change of the complex refractive index of the semiconductor resulting from the interaction with an intense pump wave has been studied extensively~\cite{Kost-book}. In contemporary literature~\cite{Lipson_review,Euser_Thesis}, it is customary to approximate the measured change in the refractive index with the free electron Drude-like model. In this case, the modification of the complex refractive index can be written as the product of the material polarizability with the optically generated FC volume density $N$, namely, as~\cite{Agrawal_Painter_review_OE_2007}
\begin{equation}\label{eq:Dn_FC}
\Delta n(x,y,z,t;\omega_s,\omega_p) = \left(\sigma_n(\omega_s;n(\omega_s)) + i \frac{c}{2 \omega_s} \sigma_a(\omega_s;n(\omega_s))\right) N(x,y,z,t;\omega_p),
\end{equation}
where the FC refractive index shift ($\sigma_n$) and the FC absorption ($\sigma_a$) cross-sections are given by
\begin{eqnarray}
\sigma_n(\omega_s;n(\omega_s)) &=& \frac{1}{2 n(\omega_s)} \frac{1}{\omega_s^2} \frac{e^2}{\epsilon_0 m_{opt}^*(\omega_s)}, \nonumber \\ \sigma_a(\omega_s;n(\omega_s)) &=& \frac{1}{n(\omega_s)} \frac{1}{c\tau_D} \frac{1}{\omega_s^2} \frac{e^2}{\epsilon_0 m_{opt}^*(\omega_s)} =  \frac{2}{c\tau_D} \sigma_n(\omega_s). \nonumber
\end{eqnarray}
Here, $\epsilon_0$ is the vacuum permittivity, $e$ is the electron charge, $c$ is the speed of light in vacuum, $m_{opt}^*$ is the effective optical mass of the FCs, and $\tau_D$ is the total FC collision rate. The cross-sections are evaluated at the signal frequency $\omega_s$, with the refractive index of the semiconductor $n$ appearing as a parameter. Note that Eq.~(\ref{eq:Dn_FC}) includes only the contribution of intraband transitions of excited FCs to the optical properties, whereas the contribution of interband transitions (such as band-filling effects, exciton formation, 
etc.) were neglected. This neglect is justified for intermediate values of FC densities that are sufficiently high to allow screening effects to wash out the excitonic signature, but that are not sufficiently high to make band-filling effects substantial~\cite{Kost-book}. This approach is valid specifically for silicon~\cite{Lipson_review} with FC densities in the order of $10^{17} - 10^{19} cm^{-3}$. In any event, the neglect of these effects will affect the following discussion only quantitatively.

The FC distribution $N$ obeys the diffusion equation,
\begin{equation}\label{eq:diffusion}
\frac{\partial N(x,y,z,t)}{\partial t} = D \nabla^2 N - \frac{N}{T_{rec}} + G(x,y,z,t;\omega_p),
\end{equation}
where $D$ is the ambipolar FC diffusion coefficient (assumed, for simplicity, to be space- and density-independent), $T_{rec}$ is the FC recombination time, and $G$ is the volume density of generated FC rate in the semiconductor. The latter equals the pump energy density dissipation rate divided by the energy of the absorbed photons
\begin{eqnarray}\label{eq:G}
G(x,y,z,t;\omega_p) &=& \frac{2 \alpha_p(\omega_p)}{\hbar \omega_p} I_p(x,y,z,t),
\end{eqnarray}
where $\alpha_p$ is the linear (amplitude) absorption coefficient of the semiconductor at the pump frequency $\omega_p$ and $I_p$ is the spatio-temporal profile of the pump field. Note that in Eq.~(\ref{eq:G}) we neglect the contribution of multiphoton absorption, an assumption valid for sufficiently weak pump intensities or for pump frequencies within the absorption band.

In the vast majority of cases, the pumping is uniform, $I_p(x,y,z,t) = I_p(t)$, and thus, diffusion is unimportant. Then, the solution of Eq.~(\ref{eq:diffusion}) is 
\begin{equation}\label{eq:N_uniform}
N(t;\omega_p) = e^{- \frac{t}{T_{rec}}} \int_{-\infty}^t e^{+\frac{t'}{T_{rec}}} G(t';\omega_p) dt',
\end{equation}
where $N_{max} = \sqrt{\pi} \frac{\alpha_p}{\hbar \omega_p} I_{p,0} T_p$, and where $I_{p,0}$ is the peak pump intensity and $T_p$ is the characteristic duration of the pump, {\XYZ which determines the rise time of the temporal step-function $m_{step}$}. In fact, if the pump duration $T_p$ is shorter than the thermalization time (the time at which the hot FCs generated by the pump start affecting the optical properties of the semiconductor), the latter should replace $T_p$. Hence, for generality, we denote the switch turn-on time as $T_{rise}$.

As noted, the typical FC recombination time is in the order of several nanoseconds~\cite{Othonos_review}; in some configurations, it can be shortened to several tens  or even to a few picoseconds~\cite{ps_Si_modulation}. Thus, since the pump duration is typically shorter than the recombination time, the FCs persist long after the pump pulse has left the system. In this case, the FC density~(\ref{eq:N_uniform}) has a fast rise time associated with the pump duration, and a much slower fall time associated with the recombination rate~\cite{Riffe_Si_e_kinetics_2002,van_Driel_kinetics_1987,Othonos_review}; accordingly, Eq.~(\ref{eq:N_uniform}) is asymptotically reduced to $N(t;\omega_p) \sim N_{max} e^{- \frac{t}{T_{rec}}} m_{step}(t/T_p)$.

More generally, the recombination time scale thus usually serves as the effective duration of the index modification and as the conventional limiting factor for ultrafast switching based on FC generation~\cite{Almeida_2004}. Specifically, the repetition rate of the pump pulses must be sufficiently low to allow the system to relax back to its equilibrium state (i.e., to allow all FCs to recombine), such that pattern effects and saturation are avoided~\cite{Tajima-theory,Tajima-exp,Lincoln-lab}.

\section{Ultrafast switching based on particle diffusion}\label{sec:diffusion-switch}

We now assume that the pump pulse generates a transient FC distribution with a one-dimensional spatial period $d$ (in the $z$ direction), such that
\begin{eqnarray}\label{eq:G_Q}
G(z,t;\omega_p) &=& \frac{\alpha_p(\omega_p)}{\hbar \omega_p} I_{p}(t) \left(1 + C\cos(2\pi z/d)\right),
\end{eqnarray}
see Fig.~\ref{fig:diffusive_switching_sims}(a). The parameter $0 < C < 1$ determines the grating contrast, with $C = 1$ corresponding to the maximal contrast. The specific form of Eq.~(\ref{eq:G_Q}) emphasizes the fact that in the majority of physical realizations of such a periodic index modification, the average of the FC pattern is not zero.

As noted previously~\cite{Sivan-Pendry-letter,Sivan-Pendry-article,Sivan-Pendry-HSM}, such an induced TBG will effectively reflect the incoming signal when the period is roughly half the wavelength of the signal wave. Thus, a TBG can be used to impose substantial amplitude modulations on the forward signal or serve as an efficient way of switching a signal between two channels, e.g., by using a circulator.

Under these conditions, FC diffusion is no longer negligible; on the contrary, it becomes the {\em dominant} effect because it causes the decay of the grating contrast. Specifically, an asymptotic solution of Eq.~(\ref{eq:diffusion}) shows that the FC distribution now has two components
\begin{eqnarray}\label{eq:N_diffusion}
N(z,t) 
&=& N_{max} \left[e^{- \frac{t}{T_{rec}}} + \cos\left(\frac{2\pi}{d}z\right) e^{- \frac{t}{T_{sw}}}\right] m_{step}(t/T_{rise}),
\end{eqnarray}
where we assume, for simplicity, that $T_{rise}$ is the shortest time scale in the system (otherwise, the relative magnitude of the two terms in Eq.~(\ref{eq:N_diffusion}) may differ slightly). The first term is associated with the uniform (average) part of the index modification, which, as described above, decays on the time scale of the recombination time $T_{rec}$. However, more importantly for our purposes, the second part of the FC density~(\ref{eq:N_diffusion}) shows that the periodic component of the generated FC density decays on a time scale given by $T_{diff} \equiv \left(\frac{d}{2\pi}\right)^2 D^{-1}$. 
The (particle) diffusion coefficient is approximately $D \sim 50cm^2/s$ for semiconductors such as silicon or germanium, and it is as much as an order of magnitude lower for doped GaAs and InP~\cite{Othonos_review}. Thus, for $d \sim \lambda_{eff}/2$ with $n_{eff} \sim 2.5$ and $\lambda_0 = 1500$ nm, we find that $T_{sw} = T_{diff} \sim 600$ fs, i.e., it is {\em orders of magnitude faster than the recombination rate}. Consequently, the effective switching time $T_{sw}$, i.e., the time at which the optical functionality (namely, the reflectivity of the system) is turned off, is determined by the diffusion time $T_{diff}$, rather than by the recombination time $T_{rec}$. Note that the high refractive index of semiconductors serves to reduce the effective wavelength, hence, to reduce the switching times. Importantly, such fast switching times are achievable with a wide range of semiconductors, with no need to resort to low temperatures, complicated nanostructuring, surface treatment, or material modifications (such as ion-implantation). As a byproduct, these fast switching times do not involve the additional scattering/loss mechanisms that are usually associated with material modifications. Notwithstanding, it is important to note that the system returns to its equilibrium state only on the time scale of $T_{rec}$. {\XYZ Thus, this approach enables switching at femtosecond {\em features}, but not at a femtosecond {\em repetition rate}, i.e., the limitation on repetition rate remains the same as in standard (i.e., uniform) switching techniques.}

On a basic physical level, our configuration exploits the non-local nature of the semiconductor nonlinearity, an aspect that is usually neglected.

\section{Ultrafast switching based on two staggered BGs - the write-erase technique}\label{sec:we-switch}

Further increase of the switching speed can be obtained by employing a second, delayed pump pulse that is spatially shifted by half the period of the generated BG. This second pump erases the BG generated by the first pump by making the FC distribution uniform in $z$; we refer to this approach as a ``write-erase'' technique. The switching time $T_{sw}$ is now simply the delay time between the two pump pulses, regardless of material parameters, such as recombination time or even diffusion times.

More quantitatively, in this case, the pumping has the form
\begin{eqnarray}\label{eq:d_eps_main_we}
G(x,y,z,t) &=& \frac{\alpha_p(\omega_p)}{\hbar \omega_p} \left\{I_p(t) \left[1 + C \cos(2\pi z/d)\right] + I_p(t - T_{sw}) \left[1 - C \cos(2\pi z/d)\right]\right\}. \nonumber
\end{eqnarray}
Accordingly, the FC density is given by
\begin{eqnarray}\label{eq:N_write-erase}
N(z,t) &=& N_{max} \{e^{- \frac{t}{T_{rec}}}  \left[m_{step}(t/T_{rise}) + m_{step}\left((t - T_{sw})/T_{rise}\right)\right] \nonumber \\
&& +\ C \cos\left(\frac{2\pi}{d}z\right) e^{- \frac{t}{T_{diff}}} \left[m_{step}(t/T_{rise}) - m_{step}\left((t - T_{sw})/T_{rise}\right)\right]\},
\end{eqnarray}
where, for simplicity, we assume $T_{diff} \gg T_{sw}$. This approach requires a delicate spatial alignment of the two pump pulses. However, the diffusion will render this approach relatively insensitive to inaccuracies in this alignment.

\section{Model - governing equations}\label{sec:model}

To exemplify in a realistic configuration the possibility of femtosecond-scale switching based on FCs as described above, let us consider a {\em pulsed} signal wave propagating in a (semiconductor) waveguide with a uniform cross-section and a refractive index of $n_{wg}$, surrounded by a medium with a refractive index $n_s$. We assume that the power of the (forward) signal wave can be represented as a product of a carrier wave $e^{- i \omega_f t + i \beta_f(\omega_f) z} + c.c.$, a modal profile $F_f(x,y;\omega_f)$ that depends on the transverse coordinates $x$ and $y$, and an envelope $A_f(z,t)$ that depends on time $t$ and the longitudinal coordinate $z$. The (potentially complex) propagation constant can be written as $\beta_f \equiv \beta_f(\omega_f) = k_f n_{eff}(\omega_f)$, where $k_f = \omega_f/c$ is the free-space wavevector and $n_{eff}$ is the effective index of the (forward) mode $f$.

We further assume that the signal wave propagation is perturbed by an intense pump beam that generates a transient, periodic modification of the refractive index of the waveguide, given by
\begin{equation}\label{eq:D_eps}
\Delta \epsilon(x,y,z,t) = 2 \epsilon_0 n_{wg} W(x,y) \Delta n(z,t),
\end{equation}
where $W(x,y)$ represents the dependence of the index change on the transverse coordinates (normalized, for simplicity, to unity peak value). This function allows for transverse FC non-uniformity due to pump absorption or reflections. {\XYZ Such inhomogeneities are small for thin waveguides and can be further reduced by adopting the detailed considerations described in~\cite{Euser_Vos_JAP_2005,Euser_Thesis}. Moreover, diffusion will serve here, again, as a fast homogenization mechanism for the FC density. Thus, we assume that realistic inhomogeneities are not sufficient to excite additional modes, especially, since those are not likely to be phase-matched. In fact, such coupling is essentially absent in thin (single mode) waveguides. }

By using Eqs.~(\ref{eq:Dn_FC}) and~(\ref{eq:N_write-erase}), we rewrite the change of the refractive index as
\begin{equation}\label{eq:D_n}
\Delta n(z,t) = \Delta n_{max} m(z,t), \quad \Delta n_{max} = \left(\sigma_n(\omega_f;n_{eff}(\omega_f)) + i \frac{c}{2 \omega_f} \sigma_a(\omega_f;n_{eff}(\omega_f))\right) N_{max},
\end{equation}
where the cross-sections are now evaluated at the effective index of the guided mode at the signal frequency ($\omega_s = \omega_f$)~\cite{Agrawal_Painter_review_OE_2007}. In that sense, it should be noted that dynamic changes of the initial values of the cross-sections and differences between the cross-sections of the forward and backward pulses due to the frequency shift are assumed to be small. Moreover, $m(z,t)$, which describes the spatio-temporal dependence of the FC density, is given by
\begin{eqnarray}\label{eq:m_general}
m(z,t) &=& m_d(t) + C \cos\left(\frac{2\pi}{d}z\right) m_c(t),
\end{eqnarray}
where $m_d$ and $m_c$ are associated with either the diffusion-based switching scheme~(\ref{eq:N_diffusion}) or with the write-erase scheme~(\ref{eq:N_write-erase}).

\begin{figure}[htbp]
  \centering{\includegraphics[scale=0.9]{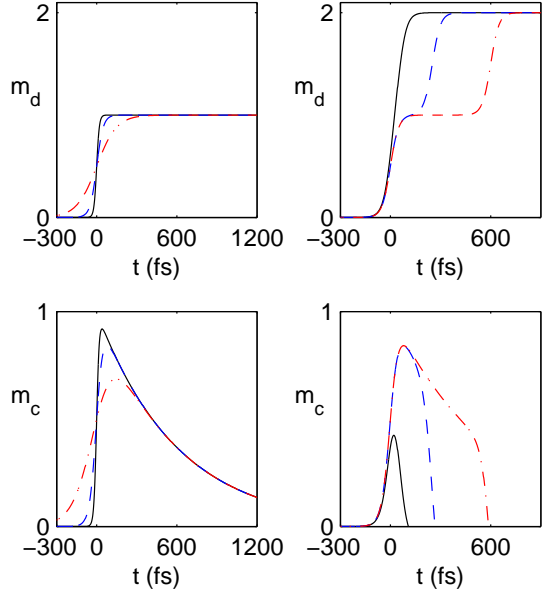}}
  \caption[]{Schematic temporal profiles of the distortion~(top) and coupling~(bottom) profiles induced in the diffusion-based switching scheme (left; varying $T_{rise}$) and write-erase switching scheme (right; varying $T_{sw}$).
   } \label{fig:m_cd}
\end{figure}

If the period of the perturbation is approximately half the effective wavelength of the signal (i.e., if the detuning of the perturbation period from the propagation constant of the (forward) mode, defined as $\delta k \equiv \beta_f - \frac{\pi}{d}$, is much smaller than $\beta_f$), then the forward signal will be coupled effectively to the backward propagating mode (whose power is given by $A_b$). In the absence of coupling to any other mode (e.g., for single mode waveguides), a straightforward extension of standard (i.e., purely spatial) coupled mode theory (see, e.g.~\cite{Yeh-book}) to pulse propagation shows that the evolution of the forward and backward pulses is given by (see~\cite{Sivan-Pendry-HSM,temporal_CMT_Bahabad}):
\begin{eqnarray}
\partial_t A_f &+& v_g \partial_z A_f(z,t) - 2 i v_g \kappa m_d(t) A_f 
= i C v_g \kappa m_c(t) e^{- 2 i \delta k z} A_b(z,t), 
\label{eq:envelope_eqs_f} \\
\partial_t A_b &-& v_g \partial_z A_b(z,t) - 2 i v_g \kappa m_d(t) A_b 
= i C v_g\kappa m_c(t) e^{2 i \delta k z} A_f(z,t), 
\label{eq:envelope_eqs_b}
\end{eqnarray}
where $v_g$ is the group velocity (equal for both modes), and the complex coupling (and distortion) coefficient is given by
\begin{equation}\label{eq:}
\kappa \equiv \epsilon_0 \omega_f \frac{\Delta n_{max}}{8} o_{fb},
\end{equation}
with
\begin{equation}\label{eq:}
o_{fb} \equiv o_{bf}^* = o_{bf} = o_{ff} = o_{bb} = \int_{-\infty}^\infty \int_{-\infty}^\infty |F_f(x,y;\omega_f)|^2 W(x,y) dx dy,
\end{equation}
being the spatial overlap of the forward ($F_f$) mode and backward mode $F_b = F_f$ with $W(x,y)$. From Eqs.~(\ref{eq:envelope_eqs_f})-(\ref{eq:envelope_eqs_b}), it is clear that $\kappa$ represents the distortion that the forward and backward waves undergo (third terms on the left hand sides) as well as the coupling between them (terms on the right hand sides). Specifically, the distortion terms arise from the fact that the mean FC distribution typically persists long after the pump pulses have left the system; see Eq.~(\ref{eq:G_Q}) and the discussion that follows it. Hence, the generated backward pulse, as well as the remains of the forward pulse, still need to propagate out of the switched region through a non-zero FC density. While doing so, they would suffer from absorption and phase-distortion (hence, potentially also shape-distortion). The distortion terms also give rise to a frequency blue-shift due to the decrease of the mean refractive index. The distortion terms have a step-function shape in time for the diffusion-based scheme, e.g., $m_d = e^{- t / T_{rec}} m_{step}$ with $m_{step} = \frac{1}{2}\left(1 + \tanh(t/T_{rise})\right)$ (see Eq.~(\ref{eq:N_diffusion})), and a double-step shape for the write-erase configuration, e.g., $m_d = e^{- \frac{t}{T_{rec}}} \left[m_{step}(t/T_{rise}) + m_{step}\left((t - T_{sw})/T_{rise}\right)\right]$ (see Eq.~(\ref{eq:N_write-erase})). On the other hand, as noted, the diffusion causes an erasure of the BG such that $m_c$ is time-localized. These temporal profiles are shown schematically in Fig.~\ref{fig:m_cd}.

The initial conditions for this set of equations are, e.g., for a Gaussian input pulse,
\begin{equation}\label{eq:ics}
A_f(z,t=0) = e^{-\left(\frac{z - z_0}{v_g T_f}\right)^2}, \quad \quad A_b(z,t = 0) = 0,
\end{equation}
where $T_f$ is the (forward) signal pulse duration and $z_0$ is its location at $t = 0$.

Equations~(\ref{eq:envelope_eqs_f})-(\ref{eq:envelope_eqs_b}) are valid for many physical configurations and are similar to those derived for bulk media in previous studies~\cite{Sivan-Pendry-letter,Sivan-Pendry-article,Sivan-Pendry-HSM}. However, the current model accounts for a realistic waveguide configuration and material parameters, and it includes additional effects, such as a possible 
non-uniformity of the pumping, absorption, complex chirp, and frequency-shift induced by the switching.

\section{Solution of the propagation equations}\label{sec:model-solution}
Eqs.~(\ref{eq:envelope_eqs_f})-(\ref{eq:envelope_eqs_b}) can be solved numerically for a wide range of parameters. They can also be solved analytically by using the Rabi approach (i.e., by ignoring the spatial relative movement of the forward and backward pulses (walk-off); see~\cite{spin-wave-reversal-karenowska-PRL}) or the low efficiency limit (see~\cite{Sivan-Pendry-letter,Sivan-Pendry-article,Sivan-Pendry-HSM}). We obtained good agreement between the numerical and analytical solutions in these limits (not shown). However, these analytical approaches hold only in first-order in $\Delta n$, and therefore, analytical predictions of forward-to-backward switching efficiencies exceeding a few tens of percent should be viewed with a grain of salt. Indeed, unlike standard coupled equations for two modes (see, e.g.,~\cite{Yariv-Yeh-book}), the current equation system cannot be solved analytically for all efficiencies due to the relative movement (walk-off) between the forward and backward waves. Thus, an accurate and complete optimization of the switching efficiency can only be accomplished numerically. Below, we present some typical examples; however, we leave the complete optimization to a future study.

The efficiency can be increased by shortening the switched region to the minimal length, which is comparable to the longitudinal length of the signal wave. In this case, the backward waves - and the remains of the forward waves - captured inside the switched region can escape quickly before suffering a considerable absorption. Typically, this escape time is short relatively to the recombination time, and, therefore, one can ignore the decay of the uniform part of the FC distribution.

In all the configurations below, we assume zero detuning, $n_{eff} = 2.5$, $\lambda = 1550$ nm, such that the diffusion time is $T_{diff} \sim 600$ fs. In addition, we set $C = 1$ and $\tau_D = 100$ fs (typical to single-crystalline silicon) and assume that the switched region has the size $B v_g T_f$ with $2 < B < 6$. Other material parameters were taken from~\cite{Si_params_van_Driel,Si_params_Linde,Euser_Thesis}.

\subsection{Diffusion-based switching scheme}

\begin{figure}[htbp]
  \centering{\includegraphics[scale=0.8]{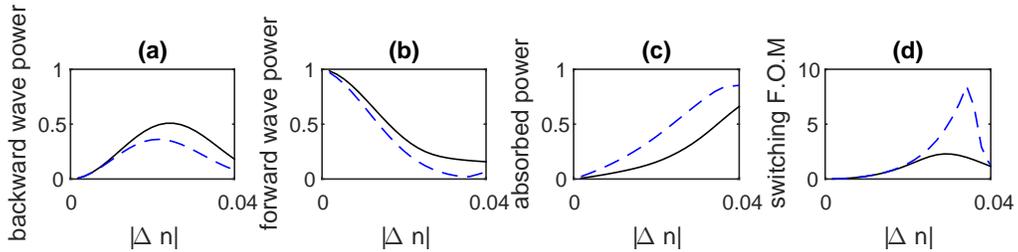}}
  \caption[]{Peak power of the (a) backward wave, (b) forward wave, (c) absorption and (d) the ratio of the backward and forward wave peak power for the diffusive switching scheme for $T_{rise} = 50$ fs, $T_f = 1000$ fs, $T_{sw} = 600$ fs. The length of the modulated region is given by  $B = 6$ (dashed blue line) and $B = 2$  (solid black line). }
  \label{fig:diffusive_switching_sims}
\end{figure}

Figure~\ref{fig:diffusive_switching_sims}(a) shows the maximal normalized power of the backward wave obtained from the numerical solution of Eqs.~(\ref{eq:envelope_eqs_f})-(\ref{eq:envelope_eqs_b}) for the diffusion-based switching scheme with $T_f = 1$ ps, $T_{sw} = T_{diff} = 600$ fs and $T_{rise} = 50$ fs. The associated changes in the refractive index are in the order of $10^{-2}$, {\XYZ which corresponds to a maximal FC density of approximately $N_{max} \sim 10^{19}/cm^3$; this is achievable in e.g., silicon with pump energy density of about $10 \mu J/cm^2$ and an absorption coefficient of $10^5$/cm ($\lambda_{pump} = 400$ nm)}. The switching efficiency initially grows for increasing switching strength, but it eventually drops as the FC absorption becomes dominant. {\XYZ The corresponding forward and absorbed peak power are shown in Figure~\ref{fig:diffusive_switching_sims}(b-c).}
For the highest index changes presented, one can see the signature of the Rabi oscillations discussed in~\cite{spin-wave-reversal-karenowska-PRL}, damped by FC absorption and distorted by forward-backward pulse walk-off. For the minimal switching spot size ($B = 2$), the reflection/switching efficiency is approximately 50\%{\XYZ. For a longer switching spot ($B = 6$), the backward pulse must propagate a longer distance through the absorbing FCs, hence, the efficiency is somewhat reduced ($\sim 36\%$). Figure~\ref{fig:diffusive_switching_sims}(d) shows the switching Figure of Merit (FOM), defined as the ratio of the backward and forward waves. One can see that the FOM reaches high values in excess of 5 for long modulation regions at experimentally accessible index modifications.}

We emphasize that the achieved switching efficiencies are unusually high for such short interaction times. Indeed, such interactions times were usually available only with the much weaker (instantaneous) Kerr effect; however, employing the stronger FC nonlinearity and relying on linear absorption enables efficient switching with relatively low pump intensities, see, e.g.,~\cite{Gaeta_PRB_2004}.

\subsection{Write-erase switching scheme}

Figure~\ref{fig:diffusive_switching_sims}(b) shows the same data as in Fig.~\ref{fig:diffusive_switching_sims}(a) for the {\XYZ write-erase scheme} for $T_f = 100$ fs. For a switching (and rise) time of $T_{sw} = T_{rise} = 50$ fs, we get approximately 35\% switching efficiency. When the delay between the pulses is extended to $T_{sw} = 150$ fs, the performance is similar but attained at correspondingly weaker pumping. Note that these efficiencies are attained with about an order of magnitude stronger pumping relative to the diffusion-based scheme, but it allows switching at features that are approximately an order of magnitude shorter. {\XYZ The FOM is substantially lower than in the diffusive-switching scheme, due to the stronger FC absorption. }

\begin{figure}[htbp]
  \centering{\includegraphics[scale=0.8]{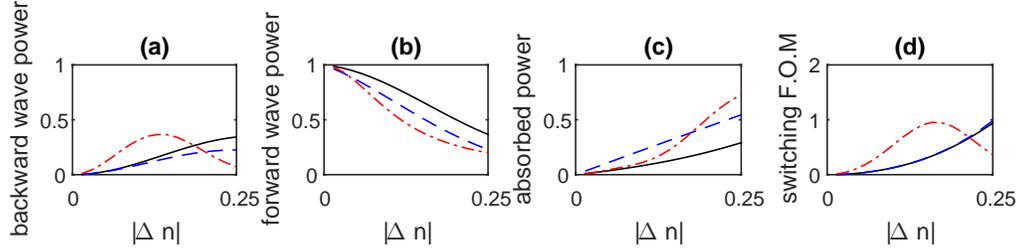}}
  \caption[]{The same as Fig.~\ref{fig:diffusive_switching_sims} for the write-erase technique with $T_{sw} = 50$ fs and $B = 2$ (black solid line) and $B = 6$ (dashed blue line) and $T_{sw} = 150$ fs and $B = 2$ (dash-dotted red line). }
  \label{fig:we_switching_sims}
\end{figure}

\subsection{Time-reversal of short optical pulses}

The switching scheme described above can be employed for wavefront-reversal of the incoming pulses, as an essential step towards a complete time-reversal of the signal pulse~\cite{Sivan-Pendry-letter,Sivan-Pendry-article,Sivan-Pendry-HSM}. This process requires the switching to be non-adiabatic, i.e., $T_{sw} \ll T_f$, to ensure that the backward pulse has the same temporal profile and spectral content as those of the incoming signal pulse; these conditions are satisfied in all the simulations shown above. However, as mentioned, the non-zero mean of the index change results in a blue-shift of the carrier frequency of the pulses. Thus, to retrieve the original carrier frequency (necessary to attain accurate time-reversal), one must shift the carrier frequency back to that of the incident pulse. This can be accomplished, for example, through a standard phase-conjugation with the pump frequency red-shifted with respect to the input frequency, such that the resulting conjugated wave will have the same frequency as that of the input pulse. Such a process can be accomplished at a high efficiency~\cite{Boyd-book} and is necessary, in any event, to complete the wavefront-reversal into a full time-reversal (see discussion in~\cite{Sivan-Pendry-article}). Alternatively, converting the backward wave frequency back to that of the incoming signal frequency can be accomplished by removing the generated FCs while the reversed pulse is still within the switched region. This alternative will serve also to reduce the absorption during the post-switching stage, thereby improving the switching efficiencies and the FOM, and will be useful to minimize pattern effects and saturation, which would accumulate after several switching cycles~\cite{Tajima-exp,Lincoln-lab,Nielsen-Mork}. For that purpose, one needs to resort to standard FC depletion techniques~\cite{LiKamWa,Feldman}.

\section{Discussion and future prospects}\label{sec:discussion}

We described a novel approach, based on FC generation, for optical switching at femtosecond time scales. By generating a periodic pattern of FCs, we enable the self-erasure of the grating by particle diffusion on sub-picosecond times. {\XYZ This is the first time, to the best of our knowledge, that diffusion is used for such a purpose}. In principle, heat diffusion, mediated, for instance, by particle collisions, will also contribute to the grating erasure. 
An even faster erasure can be achieved by using a second pump pulse that is spatially-shifted by half a period. We demonstrated the scheme for a standard silicon waveguide configuration and observed unusually high efficiencies for the pump durations in use.

As in earlier studies that rely on somewhat similar ideas {\XYZ with uniform} switching~\cite{Tajima-theory,Tajima-exp,Lincoln-lab}, our approach requires working at repetition rates that allow the system to relax to equilibrium between subsequent switching events, so as to avoid pattern effects and saturation. However, our proposed approach is more general, more compact (employing only a single channel), and spectrally selective. More importantly, our approach involves a fundamental study of time-dependent optical systems, unusual coupling between temporal and spatial degrees of freedom, and unusual wave-mixing interactions (i.e., between long and short pulses). {\XYZ We specify several applications wherein such switching characteristics are required, and elaborate on the applicability of our approach to one of those applications, namely, to time reversal of short optical pulses.}

As a final note, it should be mentioned that we assumed that the BG is longer than the pulse; however, switching with different structures, such as with BGs of finite lengths or with intentional defects, can pave the way to many novel applications. For example, such configurations may allow for pulse shaping, short pulse generation, and ultrafast amplitude modulations.

\section{Acknowledgements}
The authors acknowledge support from the NWO-nano program of the Netherlands Organization for Research (NWO), ERC Grant 279248 and a BGU M.F.S. grant. The authors would also like to acknowledge many useful discussions with N. Bar Gill, A. Ishaa'ya, R.M. de Ridder, H. Thyrrestrup, and W.L. Vos.


\begin{thebibliography}{10}

\bibitem{Eggleton_OPN}
J.~Schr\"oder, T.D. Vo, Y.~Paquot, and B.J. Eggleton.
\newblock ``Breaking the {T}bit/s barrier: Higher bandwidth optical processing,"
\newblock {Opt. Photon. News}, March, 2012.

\bibitem{Emre_OL_2013}
E.~Y\"uce, G.~Ctistis, J.~Claudon, E.~Dupuy, B.~de~Ronde, A.P. Mosk, J.-M.
  G\'erard, and W.L. Vos.
\newblock ``All-optical switching of a microcavity repeated at {TH}z rates,"
\newblock {Opt. Lett.}, ({\bf 38}):374 (2013).

\bibitem{Fan-OPN}
S.~Fan, M.~Yanik, M.~Povinelli, and S.~Sandhu.
\newblock ``Dynamic photonic crystals,"
\newblock {Opt. Phot. News}, ({\bf 18}):41 (2007).

\bibitem{Fan-reversal}
M.F. Yanik and S.~Fan.
\newblock ``Time-reversal of light with linear optics and modulators,"
\newblock {Phys. Rev. Lett.}, ({\bf 93}):173903 (2004).

\bibitem{Fan-stopping}
M.F. Yanik and S.~Fan.
\newblock ``Stopping light all optically,"
\newblock {Phys. Rev. Lett.}, ({\bf 92}):083901 (2004).

\bibitem{Longhi-reversal}
S.~Longhi.
\newblock ``Stopping and time-reversal of light in dynamic photonic structures
  via {B}loch oscillations,"
\newblock {Phys. Rev. E}, ({\bf 75}):026606 (2007).

\bibitem{Sivan-Pendry-letter}
Y.~Sivan and J.B. Pendry.
\newblock ``Time-reversal in dynamically-tuned zero-gap periodic systems,"
\newblock {Phys. Rev. Lett.}, ({\bf 106}):193902 (2011).

\bibitem{Sivan-Pendry-article}
Y.~Sivan and J.B. Pendry.
\newblock ``Theory of wave-front reversal of short pulses in dynamically-tuned
  zero-gap periodic systems,"
\newblock {Phys. Rev. A}, ({\bf 84}):033822 (2011).

\bibitem{Vos_PRB_2002}
A.F.~Koenderink P.M.~Johnson and W.L. Vos.
\newblock ``Ultrafast switching of photonic density of states in photonic
  crystals,"
\newblock {Phys. Rev. B}, ({\bf 66}):081102(R) (2002).

\bibitem{Wehrspohn_PRB_2002}
S.~Leonard, H.M. van Driel, J.~Schilling, and R.~Wehrspohn.
\newblock ``Ultrafast band-edge tuning of a 2{D} silicon photonic crystal via
  free-carrier injection,"
\newblock {Phys. Rev. B}, ({\bf 66}):161102 (2002).

\bibitem{Gaeta_PRB_2004}
V.~Almeida, C.~Barrios, R.~Panepucci, M.~Lipson, M.~Foster, D.~Ouzounov, and
  A.L. Gaeta.
\newblock ``All-optical switching on a silicon chip,"
\newblock {Opt. Lett.}, ({\bf 29}):2867--2869 (2004).

\bibitem{Euser_Vos_JAP_2005}
T.G. Euser and W.L. Vos.
\newblock ``Spatial homogeneity of optically switched semiconductor photonic
  crystals and of bulk semiconductors,"
\newblock {J. Appl. Phys.}, ({\bf 97}):043102:1--7 (2005).

\bibitem{Lipson_review}
M.~Lipson.
\newblock ``Guiding, modulating, and emitting light on silicon - challenges and
  opportunities,"
\newblock {J. Lightwave Technol.}, ({\bf 23}):4222 (2005).

\bibitem{Vos_APL_2007}
P.J. Harding, T.G. Euser, Y.~Nowicki-Bringuier, J.-M. G\'erard, and W.L. Vos.
\newblock ``Dynamical ultrafast all- optical switching of planar
  {G}a{A}s/{A}l{A}s photonic microcavities,"
\newblock {Appl. Phys. Lett.}, ({\bf 91}):111103 (2007).

\bibitem{Euser_Thesis}
T.G. Euser.
\newblock {\em Ultrafast optical switching of photonic crystals, PhD Thesis}.
\newblock 2007.

\bibitem{Othonos_review}
A.~Othonos.
\newblock ``Probing ultrafast carrier and phonon dynamics in semiconductors,"
\newblock {J. Appl. Phys.}, ({\bf 83}):1789 (1998).

\bibitem{van_Driel_kinetics_1987}
H.M. van Driel.
\newblock ``Kinetics of high-density plasmas generated in si by $1.06$- and
  $0.53$-$\mu$m picosecond laser pulses,"
\newblock {Phys. Rev. B}, ({\bf 35}):8166 (1987).

\bibitem{Riffe_Si_e_kinetics_2002}
A.J. Sabbah and D.M. Riffe.
\newblock ``Femtosecond pump-probe reflectivity study of silicon carrier
  dynamics,"
\newblock {Phys. Rev. B}, ({\bf 35}):165217 (2002).

\bibitem{Boyd-book}
R.W. Boyd.
\newblock {\em Nonlinear optics}.
\newblock (Academic Press, 2nd edition 2003).

\bibitem{Gerard_GaAs_mid_IR_relaxation}
B.~Deveaud, D.~Morris, A.~Regreny, R.~Planel, J.M. G\'erard, M.R.X. Barros, and
  P.~Becker.
\newblock ``Ultrafast relaxation of photoexcited carriers in quantum wells and
  superlattices,"
\newblock {Semicond. Scl. Technol.}, ({\bf 9}):722--726 (1994).

\bibitem{Lipson_ps_lifetime_Si_wgs}
A.~Turner-Foster, M.~Foster, J.~Levy, C.~Poitras, R.~Salem, A.~Gaeta, and
  M.~Lipson.
\newblock ``Ultrashort free-carrier lifetime in low-loss silicon nanowaveguides,"
\newblock {Opt. Express}, ({\bf 18}):3582 (2010).

\bibitem{Lipson_PINIP}
S.~Manipatruni, Q.~Xu, and M.~Lipson.
\newblock ``{PINIP} based high-speed high-extinction ratio micron-size silicon
  electrooptic modulator,"
\newblock {Opt. Express}, ({\bf 15}):13035 (2007).

\bibitem{Feldman}
J.~Feldman, E.~G\"obel, and K.~Ploog.
\newblock ``Ultrafast optical noninearities of type {II Al$_x$ Ga$_{x-1}$
  Al/AlAs} multiple quantum wells,"
\newblock {App. Phys. Lett.}, ({\bf 57}):1520 (1985).

\bibitem{LiKamWa}
P.~LiKamWa, A.~Miller, J.S. Roberts, and P.N. Robson.
\newblock ``130 ps recovery of all-optical switching in a {G}a{A}s multiquantum
  well directional coupler,"
\newblock {App. Phys. Lett.}, ({\bf 58}):2055 (1985).

\bibitem{chi_ultrafast_ion-damages_si}
F.E. Doany, D.~Grischkowsky, and C.C. Chi.
\newblock ``Carrier lifetime versus ion-implantation dose in silicon on sapphire,"
\newblock {Appl. Phys. Lett.}, ({\bf 50}):460--462 (1987).

\bibitem{Paulter_ultrafast_ion-damages_gaas}
M.B. Johnson, T.C. McGill, and N.G. Paulter.
\newblock ``Carrier lifetimes in ion-damaged {G}a{A}s,"
\newblock {Appl. Phys. Lett.}, ({\bf 54}):2424--2426 (1989).

\bibitem{ion-implant-Silberberg_1985}
Y.~Silberberg, P.W. Smith, D.A.B Miller, B.~Tell, A.C. Gossard, and
  W.~Wiegmann.
\newblock ``Fast nonlinear optical response from proton-bombarded multiple
  quantum well structures,"
\newblock {App. Phys. Lett.}, ({\bf 46}):701 (1985).

\bibitem{ps_Si_modulation}
A.~Chin, K.Y. Lee, B.C. Lin, and S.~Horng.
\newblock ``Picosecond photoresponse of carriers in {S}i ion-implanted {S}i,"
\newblock {Appl. Phys. Lett.}, ({\bf 69}):653 (1996).

\bibitem{ultrafast_H_plus_InP_APL_1991}
K.F. Lamprecht, S.~Juen, L.~Palmetshofer, and R.A. H\"opfel.
\newblock ``Ultrashort carrier lifetimes in ${H}^+$ bombarded {I}n{P},"
\newblock {Appl. Phys. Lett.}, ({\bf 59}):926--928 (1991).

\bibitem{Mourou_ultrafast_low_T}
S.~Gupta, J.~F. Whitaker, and G.A. Mourou.
\newblock ``Ultrafast carrier dynamics in {III-V} semiconductors grown by
  molecular beam epitaxy at very low substrate temperatures,"
\newblock {IEEE J. Quantum Electron.}, ({\bf 28}):2464--2472 (1992).

\bibitem{Fauchet_ultrafast_low_T}
Y.~Kostoulas, L.~Waxer, I.~Walmsley, G.~Wicks, and P.~Fauchet.
\newblock ``Femtosecond carrier dynamics in low-temperature-grown indium
  phosphide,"
\newblock {Appl. Phys. Lett.}, ({\bf 66}):1821--1823 (1995).

\bibitem{Elezzabi_ultrafast_low_T}
A.Y. Elezzabi, J.~Meyer, M.K.Y. Hughes, and S.R. Johnson.
\newblock ``Generation of 1-ps infrared pulses at 10.6 $\mu$m by use of
  low-temperature-grown {G}a{A}s as an optical semiconductor switch,"
\newblock {Opt. Lett.}, ({\bf 19}):898--900 (1994).

\bibitem{Joannopoulos-book}
J.D. Joannopoulos, S.G. Johnson, J.N. Winn, and R.D. Meade.
\newblock {\em Photonic Crystals - Molding the Flow of Light}.
\newblock (Princeton University Press, 2nd edition, 2008).

\bibitem{photosensitive-fiber-BGs-original}
S.J. Frisken.
\newblock ``Transient {B}ragg reflection gratings in erbium-doped fiber
  amplifiers,"
\newblock {Opt. Lett.}, ({\bf 17}):1776 (1992).

\bibitem{ctistis_APL_2011_Kerr}
G.~Ctistis, E.~Y\"uce, A.~Hartsuiker, J.~Claudon, M.~Bazin, J.-M. G\'erard, and
  W.L. Vos.
\newblock ``Ultimate fast optical switching of a planar microcavity in the
  telecom wavelength range,"
\newblock {Appl. Phys. Lett.}, ({\bf 98}):161114 (2011).

\bibitem{Interleaved_pns_Yu_OE_2009}
Z.-Y. Li, D.-X. Xu, W.R. McKinnon, S.~Janz, J.H. Schmid, P.~Cheben, and J.~Yu.
\newblock ``Silicon waveguide modulator based on carrier depletion in
  periodically interleaved pn junctions,"
\newblock {Optics Express}, ({\bf 17}):15947--15958 (2009).

\bibitem{Interleaved_pns_Yu_OE_2012}
X.~Xiao, H.~Xu, X.~Li, Y.~Hu, K.~Xiong, Z.~Li, T.~Chu, Y.~Yu, and J.~Yu.
\newblock ``25 {G}bit/s silicon microring modulator based on
  misalignment-tolerant interleaved pn junctions,"
\newblock {Optics Express}, ({\bf 20}):2507--2515 (2012).

\bibitem{Yeh-book}
P.~Yeh.
\newblock {\em Optical Waves in Layered Media}.
\newblock (Wiley-Interscience, 2nd edition, 2005).

\bibitem{Kost-book}
E.~Garmire and A.~Kost.
\newblock {\em Nonlinear optics in Semiconductors}.
\newblock (Academic Press, 1999).

\bibitem{Agrawal_Painter_review_OE_2007}
Q.~Lin, O.J. Painter, and G.P. Agrawal.
\newblock ``Nonlinear optical phenomena in silicon waveguides: modeling and
  applications,"
\newblock {Opt. Express}, ({\bf 15}):16604 (2007).

\bibitem{Almeida_2004}
V.~Almeida, C.~Barrios, R.~Panepucci, and M.~Lipson.
\newblock ``All-optical control of light on a silicon chip,"
\newblock {Nature}, ({\bf 431}):1081 (2004).

\bibitem{Tajima-theory}
K.~Tajima.
\newblock ``All-optical switch with switch-off time unrestricted by carrier
  lifetime,"
\newblock {Jpn. J. Appl. Phys.}, ({\bf 32}):L1746--L1749 (1993).

\bibitem{Tajima-exp}
S.~Nakamura, Y.~Ueno, K.~Tajima, J.~Sasaki, T.~Sugimoto, T.~Kato, T.~Shimoda,
  M.~Itoh, H.~Hatakeyama, T.~Tamanuki, and T.~Sasaki.
\newblock ``Demultiplexing of 168-{G}b/s data pulses with a hybrid-integrated
  symmetric {M}ach–{Z}ehnder all-optical switch,"
\newblock {IEEE Photon. Tch. Lett.}, ({\bf 12}):425 (2000).

\bibitem{Lincoln-lab}
N.S. Patel, K.L. Hall, and K.A. Rauschenbach.
\newblock ``Interferometric all-optical switches for ultrafast signal processing,"
\newblock {Appl. Optics}, ({\bf 37}):2831 (1998).

\bibitem{Sivan-Pendry-HSM}
Y.~Sivan and J.B. Pendry.
\newblock ``Broadband time-reversal of optical pulses using a switchable
  photonic-crystal mirror,"
\newblock {Opt. Express}, ({\bf 19}):14502 (2011).

\bibitem{temporal_CMT_Bahabad}
B.~Dana, L.~Lobachinsky, and A.~Bahabad.
\newblock ``Spatiotemporal coupled-mode theory in dispersive media under a
  dynamic modulation,"
\newblock {Opt. Comm.}, ({\bf 324}):165--167 (2014).

\bibitem{spin-wave-reversal-karenowska-PRL}
A.~Karenowska, J.~Gregg, V.~Tiberkevich, A.~Slavin, A.~Chumak, A.~Serga, and
  B.~Hillebrands.
\newblock ``Oscillatory energy exchange between waves coupled by a dynamic
  artificial crystal,"
\newblock {Phys. Rev. Lett.}, ({\bf 108}):015505 (2012).

\bibitem{Yariv-Yeh-book}
A.~Yariv and P.~Yeh.
\newblock {\em Photonics}.
\newblock (Oxford University Press, 6th edition, 2007).

\bibitem{Si_params_van_Driel}
H.M. van Driel.
\newblock ``Optical effective mass of high density carriers in silicon,"
\newblock {Appl. Phys. Lett.}, ({\bf 44}):617 (1984).

\bibitem{Si_params_Linde}
K.~Sokolowski-Tinten and D.~von~der Linde.
\newblock ``Generation of dense electron-hole plasmas in silicon,"
\newblock {Phys. Rev. B}, ({\bf 61}):2643 (1984).

\bibitem{Nielsen-Mork}
M.L. Nielsen and J.~Mork.
\newblock ``Increasing the modulation bandwidth of
  semiconductor-optical-amplifier-based switches by using optical filtering,"
\newblock {J. Opt. Soc. Am. B}, ({\bf 21}):1606 (2004).

\end{thebibliography}
\end{document}